\documentclass[twocolumn,prb,amsmath,amssymb]{revtex4}
\usepackage{graphicx}
\usepackage{booktabs}
\usepackage{multirow}
\usepackage{amssymb}

\renewcommand{\vec}[1]{\mathrm{\mathbf{#1}}}

\begin{document}

\title{Graphene Photonics and Optoelectronics}
\author{F. Bonaccorso, Z. Sun, T. Hasan, A. C. Ferrari}
\email{acf26@eng.cam.ac.uk}
\affiliation{Department of Engineering, University of Cambridge, Cambridge CB3 0FA, UK}

\begin{abstract}
The richness of optical and electronic properties of graphene attracts enormous interest. Graphene has high mobility and optical transparency, in addition to flexibility, robustness and environmental stability. So far, the main focus has been on fundamental physics and electronic devices. However, we believe its true potential to be in photonics and optoelectronics, where \textit{the combination} of its unique optical and electronic properties can be fully exploited, even in the absence of a bandgap, and the linear dispersion of the Dirac electrons enables ultra-wide-band tunability. The rise of graphene in photonics and optoelectronics is shown by several recent results, ranging from solar cells and light emitting devices, to touch screens, photodetectors and ultrafast lasers. Here we review the state of the art in this emerging field.
\end{abstract}
\maketitle

\section{\label{In}Introduction}
Electrons propagating through the bi-dimensional structure of graphene behave as massless Dirac fermions\cite{Geim_nmat,Charlier,Wallace_1947}, having a linear energy-momentum relation. Consequently, graphene exhibits electronic properties for a two dimensional (2d) gas of charged particles described by the relativistic Dirac equation, rather than the non-relativistic Schr\"odinger equation with an effective mass\cite{Geim_nmat,Charlier}, with carriers mimicking particles with zero mass and an effective ``speed of light''$\sim 10^6$m/s.

Graphene has revealed a variety of transport phenomena characteristic of 2d Dirac fermions, such as specific integer and fractional quantum Hall effects\cite{Zhang_Nature2005,andrei}, a "minimum" conductivity$\sim 4e^2/h$ even when the carrier concentration tends to zero\cite{Geim_nmat}, Shubnikov-de Haas oscillations with phase shift $\pi$ due to Berry's phase\cite{Geim_nmat}. Mobilities, $\mu$, up to $10^{6}cm^{2}/Vs$ are observed in suspended samples. This, combined with near-ballistic transport at room temperature, makes graphene a potential material for nanoelectronics\cite{Lemme_IEEE,Han_PRL}, especially for high frequency\cite{lin_10}.

Graphene also shows remarkable optical properties. Despite being a single atom thick, it can be optically visualized\cite{Casi_nanolett,Blake_APL}. Its transmittance can be expressed in terms of the fine structure constant\cite{Nair_Science,Kravets}. The linear dispersion of the Dirac electrons enables broadband applications. Saturable absorption is observed as a consequence of Pauli blocking\cite{Hasan_AM,Sun_graphene}, non equilibrium carriers result in hot luminescence\cite{NLPL}. Chemical and physical treatments also enable photoluminescence\cite{gokusacsnano,edaadvmater,sunnanores,luoapl}. These properties make it an ideal photonic and optoelectronic material.

\section{\label{EOR}Electronic and optical properties}
\subsection{\label{EP}Electronic properties}
The electronic structure of single layer graphene (SLG) can be described using a tight-binding Hamiltonian\cite{Charlier,Wallace_1947}. Since the bonding and anti-bonding $\sigma$ bands are well separated in energy ($>$10 eV at $\Gamma$), they can be neglected in semi-empirical calculations, retaining only the two remaining $\pi$ bands\cite{Wallace_1947}. The electronic wavefunctions from different atoms on the hexagonal lattice overlap. However, such an overlap between the $p_z$($\pi$) and the $s$ or $p_x$ and $p_y$ orbitals is strictly zero by symmetry. Consequently, the $p_z$ electrons, which form the $\pi$ bonds, can be treated independently from the other valence electrons. Within this $\pi$-band approximation:
\begin{eqnarray}
E^{\pm}(k_{x},k_{y}) =  \pm \gamma_{0}
\sqrt{1 + 4\cos\frac{\sqrt{3}k_{x}a}{2} \cos\frac{k_{y}a}{2}+ 4\cos^{2} \frac{k_{y}a}{2}}\nonumber\\
\label{Egraphene}
\end{eqnarray}
\noindent where $a=\sqrt{3}a_{\rm CC}$ ($a_{\rm CC}$=1.42~\AA\ is the carbon-carbon distance) and $\gamma_{0}$ is the transfer integral between first neighbor $\pi$ orbitals (typical values for $\gamma_0$ are 2.9-3.1eV). The ${\bf k}=(k_{x},k_{y})$ vectors in the first Brillouin zone (BZ) constitute the ensemble of available electronic momenta.

With one {\it p}$_z$ electron per atom in the $\pi$-$\pi^{\ast}$ model (the three other ${\it s, p_x, p_y}$ electrons fill the low-lying $\sigma$ band), the (-) band (negative energy branch) in Eq.~\ref{Egraphene} is fully occupied, while the (+) branch is totally empty. These occupied and unoccupied bands touch at the $K$ points. The Fermi level $E_F$ is the zero-energy reference, and the Fermi surface is defined by $K$ and $K'$.

Expanding Eq.~\ref{Egraphene} at $K(K')$ yields the linear $\pi$ and $\pi^{\ast}$ bands for Dirac fermions:
\begin{eqnarray}
E^{\pm}(\boldsymbol{\kappa}) =  \pm \hbar v_F|\boldsymbol{\kappa}|
\label{Edirac}
\end{eqnarray}
where $\boldsymbol{\kappa=k-K}$, and $v_F$ is the electronic group velocity: $v_F=\sqrt{3}\gamma_0 a/2\hbar \sim 1\times 10^6 m/s$.

The linear dispersion given by Eq.~\ref{Edirac} is the solution to the following effective Hamiltonian at the $K(K')$ point:
\begin{eqnarray}
\boldsymbol{H} = \hbar v_F(\boldsymbol{\sigma}\cdot\boldsymbol{
\kappa})\label{hamiltonian}
\end{eqnarray}
where $\boldsymbol{\kappa}=-i\nabla$, and $\boldsymbol{\sigma}$'s are the pseudo-spin Pauli matrices operating in the space of the electron amplitude on the A-B sublattices of graphene.

\subsection{\label{ab}Linear optical absorption}
\begin{figure}
 \centerline{\includegraphics[width=60mm]{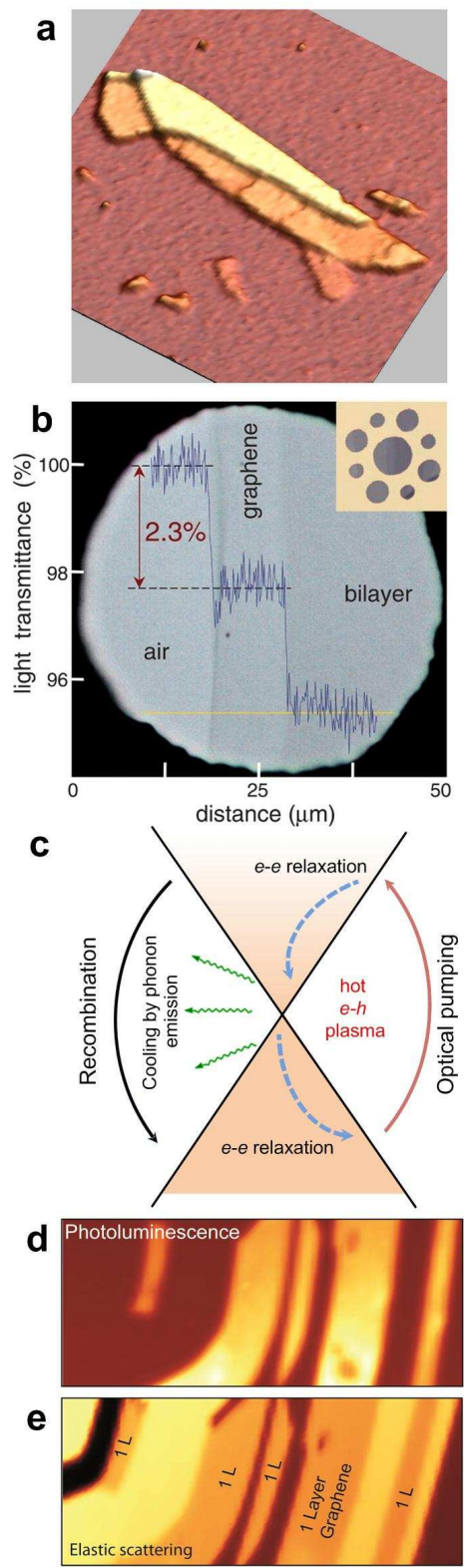}}
 \caption{\label{Figure_1}(a) Elastic light scattering (Rayleigh) image of a flake with varying number of layers (adapted from Ref. \onlinecite{Casi_nanolett}).(b) Transmittance for increasing number of layers (adapted from Ref. \onlinecite{Nair_Science}).(c) Schematic of photoexcited electron kinetics in graphene, with possible relaxation mechanisms for the non-equilibrium electron population.(d) PL;(e) elastic scattering images of an oxygen treated flake (adapted from Ref. \onlinecite{gokusacsnano}).}
\end{figure}

The optical image contrast can be used to identify graphene on top of a Si/SiO$_2$ substrate (Fig.\ref{Figure_1}a). This scales with the number of layers and is the result of interference, with SiO$_2$ acting as a spacer. The contrast can be maximized by adjusting the spacer thickness or the light wavelength\cite{Casi_nanolett,Blake_APL}.

The transmittance (T) of a freestanding SLG can be derived by applying the Fresnel equations in the thin film limit for a material with a fixed universal optical conductance\cite{Kuzmenko_2008_PRL} $G_0=\frac{e^{2}}{4 \hbar}\sim6.08\times10^{-5}\Omega^{-1}$, to give:
 \begin{equation} \label{equ_8}
T  = (1+ 0.5\pi\alpha)^{-2} \approx 1-\pi\alpha \approx 97.7 \%
\end{equation}
where $\alpha = \frac{e^{2}}{4\pi\epsilon_0\hbar c}=\frac{G_0}{ \pi\epsilon_0 c}\approx 1/137$ is the fine structure constant\cite{Nair_Science}. Graphene only reflects $<0.1\%$ of the incident light in the visible region\cite{Nair_Science}, raising to $\sim$2\% for 10 layers\cite{Casi_nanolett}. Thus, we can take the optical absorption of graphene layers to be proportional to the number of layers, each absorbing A=1-T=$\pi\alpha$=2.3\% over the visible spectrum\cite{Nair_Science}, Fig.\ref{Figure_1}b. In a few layer graphene (FLG) sample, each sheet can be seen as a bi-dimensional electron gas, with little perturbation from the adjacent layers, making it optically equivalent to a superposition of almost non-interacting SLG\cite{Casi_nanolett}. The absorption of SLG is quite flat from 300 to 2500nm with a peak in the UV region ($\sim$250nm), attributed to inter-band electronic transition from the unoccupied $\pi^{*}$ states\cite{Kravets}. In FLG, other absorption features can be seen at lower energies, associated with inter-band transitions\cite{Wang_science_2008,Mak}.

\subsection{\label{sat_ab}Saturable absorption}
Interband excitation by ultrafast optical pulses produces a non-equilibrium carrier population in the valence and conduction bands, Fig.\ref{Figure_1}c. In time-resolved experiments\cite{Breusing_PRL} two relaxation time scales are typically seen. A faster one, $\sim{100}fs$, usually associated with carrier-carrier intraband collisions and phonon emission, and a slower one, on a ps scale, corresponding to electron interband relaxation and cooling of hot phonons\cite{kampfrathprl05,lazzeriPRL}.

The linear dispersion of the Dirac electrons implies that for any excitation there will always be an electron-hole pair in resonance. A quantitative treatment of the electron-hole dynamics requires the solution of the kinetic equation for the electron and hole distribution functions $f_e(\vec{p})$ and $f_h(\vec{p})$, $\vec{p}$~being the momentum counted from the Dirac point\cite{Sun_graphene}. If the relaxation times are shorter than the pulse duration, during the pulse the electrons reach a stationary state and collisions put electrons and holes in thermal equilibrium at an effective temperature\cite{Sun_graphene}. The populations determine electron and hole densities, total energy density and a reduction of photon absorption per layer, due to Pauli blocking, by a factor of $\Delta A/A=[1-f_e(\vec{p})][1-f_h(\vec{p})]-1$. Assuming efficient carrier-carrier relaxation (both intraband and interband) and efficient cooling of the graphene phonons, the main bottleneck is energy transfer from electrons to phonons\cite{Sun_graphene}.

For linear dispersions near the Dirac point, pair carrier collisions cannot lead to interband relaxation, thereby conserving the total number of electrons and holes separately\cite{Sun_graphene,Guinea96}. Interband relaxation by phonon emission can occur only if the electron and hole energies are close to the Dirac point (within the phonon energy). Non-equilibrium electron-hole recombination is also possible\cite{NLPL}. For graphite flakes the situation is different: the dispersion is quadratic, and pair carrier collisions can lead to interband relaxation. Thus, in principle, decoupled SLG can provide the highest saturable absorption for a given amount of material\cite{Sun_graphene}.

\subsection{\label{PL}Luminescence}
Graphene could be made luminescent by inducing a band gap, following two main routes. One is by cutting it into ribbons and quantum dots, the other is by chemical or physical treatments, to reduce the connectivity of the $\pi$ electrons network. Even though graphene nanoribbons (GNRs) have been produced, with varying band gaps\cite{Han_PRL}, to date no photoluminescence (PL) has been reported from them. However, bulk graphene oxide (GO) dispersions and solids do show a broad PL\cite{edaadvmater,sunnanores,luoapl,KPlohACSnano}. Individual graphene flakes can be made brightly luminescent by mild oxygen plasma treatment\cite{gokusacsnano}. The resulting PL is uniform across large areas, as shown in Figs. \ref{Figure_1}d,e, where a PL map and the corresponding elastic scattering image are compared. It is possible to make hybrid structures by etching just the top layer, while leaving underlying layers intact\cite{gokusacsnano}. This combination of photoluminescent and conductive layers could be used in sandwich light-emitting diodes. Luminescent graphene-based materials can now be routinely produced covering the infrared (IR), visible and blue spectral range\cite{gokusacsnano,edaadvmater,sunnanores,luoapl,KPlohACSnano}.

Even though some groups assigned PL in GO to band-gap emission from electron confined $sp^{2}$ islands\cite{edaadvmater,sunnanores,luoapl}, this is more likely to arise from oxygen related defect states\cite{gokusacsnano}.  Whatever the origin, fluorescent organic compounds are of significant importance to the development of low-cost opto-electronic devices\cite{sheaths}. Blue PL from aromatic or olefinic molecules is particularly important for display and lighting applications\cite{rothberg}. Luminescent quantum dots are widely used for bio-labeling and bio-imaging. However, their toxicity and potential environmental hazard limit widespread use and \textit{in-vivo} applications. Fluorescent bio-compatible carbon-based nanomaterials, may be a more suitable alternative. Fluorescent species in the IR and near-IR are useful for biological applications since cells and tissues exhibit little auto-fluorescence in this region\cite{Frangioni_JV}. Ref. \onlinecite{sunnanores} exploited PL GO for live cell imaging in the near-infrared with little background.

Ref. \onlinecite{Wang_science_2008} reported a gate-controlled, tunable gap up to 250meV in BLG. This may enable novel photonic devices for far IR light generation, amplification and detection.

Broadband non-linear PL is also possible following non equilibrium excitation of untreated graphene layers, Fig 1c, as recently reported by several groups\cite{NLPL}. Emission was observed throughout the visible spectrum, for energies both higher and lower than the exciting one, in contrast to conventional PL processes\cite{NLPL}. This broadband non linear-PL arises from recombination of a distribution of non-equilibrium electrons and holes, generated by rapid scattering between photo-excited carriers after optical excitation\cite{NLPL}. It scales with the number of layers and can be used as a quantitative imaging tool, as well as to reveal the hot electron-hole plasma dynamics\cite{NLPL}, Fig 1c.

Electroluminescence was also recently reported in pristine graphene\cite{Essig}. Although the power conversion efficiency is lower than carbon nanotubes (CNTs), this could lead to novel emitting devices based entirely on graphene.

\section{\label{FT} Production}
Graphene was first produced by micromechanical exfoliation of graphite\cite{Novoselov_PNAS}. This still gives the best samples in terms of purity, defects, mobility and optoelectronics properties. However, large scale assembly is needed for the widespread application of this material. Several techniques have been developed to provide a steady supply of graphene in large areas and quantities, amenable for mass applications. These comprise growth by chemical vapor deposition\cite{Karu_JAP_1966,Obraztsov_Carbon,Kim_nature,Reina_NL_2009,Bae_arxiv}, segregation by heat treatment of carbon containing substrates\cite{Berger,Sutter_nmat,Emtsev_nmat}, liquid phase exfoliation\cite{Li_Science,Hernandez_nnano,Lotya_JACS,Valles_Jacs,Stankovich_Nature}.
In fact, most of these methods date back several decades. The current interest in graphene pushed these early approaches to large yields, controlled growth, large areas, and enabled in just 6 years to go from micrometer flakes to near mass production of layer controlled samples.

\subsection{\label{MC} Micromechanical cleavage} Micromechanical cleavage (MC)\cite{Novoselov_PNAS} consists in peeling off a piece of graphite by means of an adhesive tape. MC has been optimized to give SLG up to mm in size, of high structural and electronic quality. Although this is the method of choice for fundamental research, and most key results on individual SLG were obtained on MC flakes, it has disadvantages in terms of yield and throughput, and is impractical for large scale applications.

\subsection{\label{LPE} Liquid phase exfoliation} Liquid phase exfoliation (LPE) of graphite consists in chemical wet dispersion followed by ultrasonication, both in aqueous\cite{Lotya_JACS} and non-aqueous solvents\cite{Hernandez_nnano}. Up to$\sim70\%$ SLG can be achieved by mild sonication in water with sodium deoxycholate followed by sedimentation based-ultracentrifugation\cite{Marago}.

Bile salts surfactants also allow the isolation of flakes with controlled thickness, when combined with density gradient ultracentrifugation (DGU)\cite{Green_NL}. Exfoliation of graphite intercalated compounds\cite{Valles_Jacs} and expandable graphite\cite{Dai_nnano} was also reported.

LPE can also give GNRs with widths$<$10nm\cite{Li_Science} and offers advantages of scalability and no requirement of expensive growth substrates. Furthermore it is an ideal means to produce films and composites.

\subsubsection{\label{GO}Graphene oxide}
LPE of graphite was first obtained through sonication of graphite oxide\cite{Stankovich_Nature}, following the 60 years old Hummers method\cite{Hummers_1958}, to produce graphene oxide (GO). The oxidation of graphite in the presence of acids and oxidants, proposed already in the nineteenth century\cite{Brodie,Staudenmaier}, disrupts the sp$^{2}$-network and introduces hydroxyl or epoxide groups\cite{Mattevi_AFM,cai_sci_08} with carboxylic or carbonyl groups attached to the edges. These make GO sheets readily dispersible in water and several other solvents. Although large GO flakes can be produced, these are intrinsically defective, and electrically insulating. Despite several attempts\cite{Stankovich_Nature,Mattevi_AFM}, reduced GO (RGO) does not fully regain the pristine graphene electrical conductivity\cite{Mattevi_AFM,Eda_nnano}. It is thus important to distinguish between dispersion processed graphene flakes, retaining the electronic properties of graphene, and insulating GO layers.

\subsection{\label{ESic} Chemical vapor deposition}
FLGs were grown more than 40 years ago by Chemical Vapor Deposition (CVD)\cite{Karu_JAP_1966}. SLG and FLG can now be produced on various substrates by feeding hydrocarbons at a suitable temperature\cite{Karu_JAP_1966,Obraztsov_Carbon,Kim_nature,Reina_NL_2009,Bae_arxiv,Oshima_1997,Lband_SS,Wang_Carbon}. The scale of progress in CVD growth is given by Ref. \onlinecite{Bae_arxiv}, where samples over 60cm were achieved. Plasma-enhanced CVD can be applied on substrates without catalyst\cite{Wang_Carbon}. Note that most as-grown CVD samples are multilayer. Even if their Raman spectrum appears similar\cite{Kim_nature,Reina_NL_2009} to the ideal SLG one\cite{Ferrari_PRL}, this is just an indication of electronic decoupling of the layers, not a definite proof of SLG growth.

\subsection{\label{EG}Carbon segregation}
Carbon segregation from silicon carbide\cite{Berger,Acheson,Badami_62} or metal substrates\cite{Sutter_nmat,Oshima_1997,Isett_1976,Gamo_SS_1997,Rosei_PRB_83,Ueta_SS,Marchini_PRB_2007}, following high temperature annealing, can produce graphene. As early as 1896 E. G. Acheson reported a method to produce graphite from SiC\cite{Acheson}, while the segregation of graphene from Ni(111) was investigated over 30 years ago\cite{Isett_1976}. High quality layers can now be produced on SiC in an Ar atmosphere\cite{Emtsev_nmat} and electronic decoupling from the underlying SiC substrate can be achieved by H treatment\cite{Stark}.

\subsection{\label{CS}Chemical synthesis}
Graphene or carbon nanosheets can also be chemically synthesized\cite{Chouchair_nnano}. Total organic synthesis yields graphene-like poly-aromatic hydrocarbons (PAHs)\cite{Wang_Angwandte}. These synthetic nanographenes (NGs) can then be assembled to form larger layers. Supramolecular interactions can be used to cover SLG with PAHs, keeping the $sp^{2}$ network intact, without compromising the transport properties. NGs form ordered layers, with precise control of orientation and spacing\cite{Mullen_CR}. These interact with the graphene backbone allowing in principle to control and tune its optoelectronic properties\cite{Mullen_CR}.

\subsection{Deterministic placement}
A fundamental step to produce useful devices is the deterministic placement of graphene on pre-defined positions on a substrate of choice. Transfer processes are common in the semiconductor industry. Extensive experience of transfer was developed for carbon nanotubes (CNTs). Ref. \onlinecite{Reina2008} reported transfer of SLG and FLG from SiO$_{2}$/Si to other substrates. A layer of poly(methyl methacrylate) (PMMA) was coated on graphene deposited on SiO$_{2}$, subsequently detached by partial SiO$_{2}$ etch\cite{Reina2008}. The PMMA/graphene membrane was then placed over the target substrate and PMMA dissolved with acetone\cite{Reina2008}. Ref. \onlinecite{Kim_nature} used a dry-method based on a polydimethylsiloxane (PDMS) stamp to transfer patterned films. Ref. \onlinecite{Bae_arxiv} scaled the process to a roll-based layer-by-layer transfer onto plastic substrates.

We developed a procedure for deterministic placement, following transfer. This exploits a water layer between the PMMA/graphene foil and the substrate, enabling the PMMA to move. This allows to place graphene layers on any substrate in any predefined location, prepare "artificial" multi-layers, create sandwich structures with other materials (such as BN, MoS$_2$), etc. We will show an example of this technique in Section \ref{SA}, by placing graphene on the core of an optical fibre.

Large scale placement of LPE samples can be achieved by spin coating and Langmuir-Blogdett\cite{Dai_nnano}, even though with lack of positional control. Surface modification by self-assembled monolayers (SAMs) can enable targeted deposition of graphene flakes in a large scale. Di-electrophoresis allows controlled placement of individual graphene flakes between pre-patterned electrodes\cite{krupke}. Inkjet printing is another attractive technique\cite{Beecher_JAP}, and could directly "write" optoelectronic devices.

\section{\label{AP} Photonics and Optoelectronics applications}
\begin{figure*}
\centerline{\includegraphics[width=170mm]{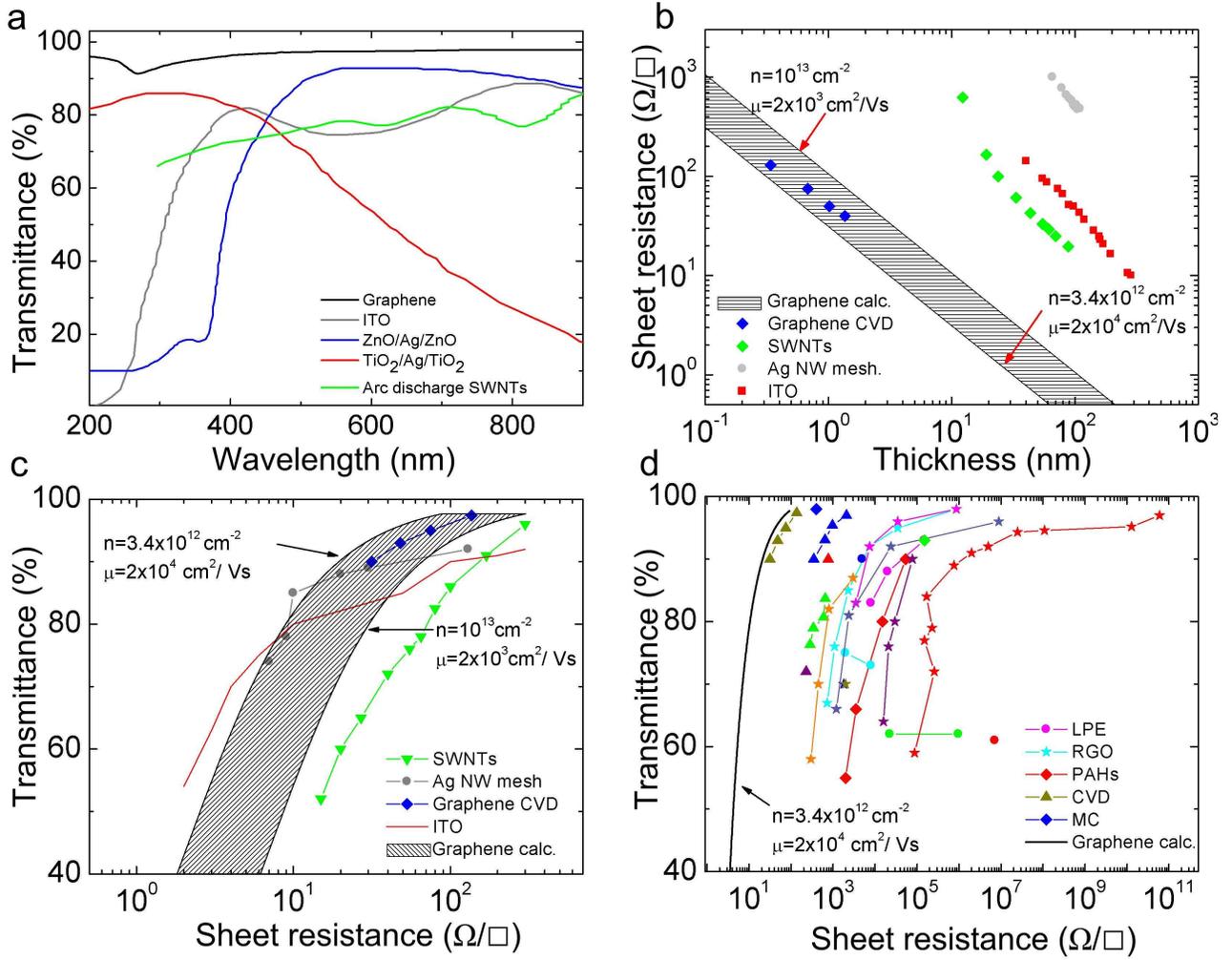}}
\caption{\label{Figure_2}a) T for different transparent conductors: GTCFs\cite{Bae_arxiv}, single wall carbon nanotubes(SWNTs)\cite{Geng_JACS}, ITO\cite{Peumans_NL}, ZnO-Ag-ZnO\cite{Sahu_ASS}, TiO$_2$/Ag/TiO$_2$\cite{Hamberg_JAP_1986}, where Ag is Silver and TiO$_2$ is Titanium dioxide. b) Thickness dependence of $R_s$. blue rhombuses, roll to roll GTCFs\cite{Bae_arxiv}; red squares, ITO\cite {Peumans_NL}; grey dots, metal wires\cite{Peumans_NL}; green rhombuses, SWNT\cite{Geng_JACS}. Two limiting lines for GTCFs are also plotted (hatched area), calculated from Eq. 8 using typical values for n and $\mu$. c) T vs $R_s$ for different transparent conductors: blue rhombuses, roll to roll GTCFs\cite{Bae_arxiv}; red line, ITO\cite {Peumans_NL}; grey dot, metal wires\cite{Peumans_NL}; green triangles, SWNTs\cite{Geng_JACS}. Hatched area, limiting lines for GTCFs calculated using n and $\mu$ as in b). d) T vs $R_{s}$ for GTCFs grouped according to production strategies: CVD\cite{Kim_nature,Reina_NL_2009,Bae_arxiv,De_Arco}, micro-mechanical cleavage (MC)\cite{Blake_nanoletters}, organic synthesis\cite{Wang_Angwandte}, LPE of pristine graphene\cite{Hernandez_nnano,Lotya_JACS,Green_NL,Blake_nanoletters} or GO\cite{Mattevi_AFM,Eda_nnano,Becerril_ACSnano,Wu_APL,Wang_NL}. A theoretical line as for Eq. 8 is also plotted for comparison.}
\end{figure*}
\subsection{\label{TC} Transparent conductors}
Optoelectronic devices such as displays, touch-screens, light emitting diodes, solar cells, require materials with low sheet resistance $R_{s}$ and high transparency. In a thin-film $R_{s}=\rho$/d, where d is the film thickness and $\rho=1/\sigma_{dc}$ the resistivity. For a rectangle of length L and width W, the resistance R is:
\begin{equation} \label{equ_6}
R = \frac{\rho}{d}\times\frac{L}{W}=R_{s}\times\frac{L}{W}
\end{equation}

The term L/W can be seen as the number of squares of side W that can be superimposed on the resistor without overlapping. Thus, even if R$_{s}$ has units of ohms as R, it is historically quoted in "ohms per square" ($\Omega/\Box$).

Current transparent conductors (TC) are semiconductor-based\cite{Hamberg_JAP_1986}: doped Indium Oxide (In$_2$O$_3$)\cite{Holland_53}, Zinc Oxide (ZnO)\cite{Minami_SST_2005}, Tin Oxide (SnO$_2$)\cite{Hamberg_JAP_1986}, as well as ternary compounds based on their combinations\cite{Hamberg_JAP_1986,Minami_SST_2005, Granqvist_07}. The dominant material is indium tin-oxide (ITO), a doped n-type semiconductor composed of$\sim90\%$ In$_2$O$_3$, and$\sim10\%$ SnO$_2$\cite{Hamberg_JAP_1986}. The electrical and optical properties of ITO are strongly affected by the impurities\cite{Hamberg_JAP_1986}. Sn atoms act as n-type donors\cite{Hamberg_JAP_1986}. ITO has strong absorption above 4eV due to interband transitions\cite{Hamberg_JAP_1986}, with other features at lower energy related to scattering of free electrons with Sn atoms or grain boundaries\cite{Hamberg_JAP_1986}. ITO is commercially available with T$\sim 80\%$ and $R_{s}$ as low as $10 \Omega/\Box$ on glass\cite{Minami_SST_2005}, and $\sim60-300\Omega/\Box$  on polyethylene terephthalate (PET)\cite{Granqvist_07}. Note that T is typically quoted at 550nm, since this is where the human eye spectral response is maximum\cite{Hamberg_JAP_1986}.

ITO suffers severe limitations: an ever increasing cost due to Indium scarcity\cite{Hamberg_JAP_1986}, processing requirements, difficulties in patterning\cite{Granqvist_07,Hamberg_JAP_1986}, sensitivity to acidic and basic environments. Moreover, ITO is brittle and can easily wear out or crack when used in applications where bending is involved, such as touch screens and flexible displays\cite{sheraw_02}. This demands new TC materials with improved performance. Metal grids\cite{Peumans_NL}, metallic nanowires\cite{De_acsnano2}, or other metal oxides\cite{Granqvist_07} have been explored as alternative. Nanotubes and graphene also show great promise. In particular, graphene films have higher T over a wider wavelength range with respect to SWNT films\cite{Geng_JACS, Wu_NT_Science,Coleman_ACS}, thin metallic films\cite{Peumans_NL,De_acsnano2}, and ITO\cite{Hamberg_JAP_1986,Minami_SST_2005}, Fig. \ref{Figure_2}a.

We now give a relation between T and R$_s$ for graphene films of varying doping. From Eq. 4, T depends on the optical conductivity G$_0$,
\begin{equation} \label{equ_T}
T =\left(1+ \frac{G_0}{2 \epsilon_0 c}N\right)^{-2},
\end{equation}
where N is the number of layers. R$_{s}$ is linked to the bi-dimensional dc conductivity $\sigma_{2d,dc}$ by:
\begin{equation} \label{equ_Rs}
R_{s} = (\sigma_{2d,dc} N)^{-1}
\end{equation}
Combining Eqs. \ref{equ_T},\ref{equ_Rs}, eliminating N, gives:
\begin{equation} \label{equ_13}
T = \left(1+ \frac{Z_{0}}{2R_{s}} \frac{G_0}{\sigma_{2d,dc}}\right)^{-2}
\end{equation}
where, $Z_{0}={\frac{1}{\epsilon_0 c}}$=377$\Omega$ is the free space impedance, $\epsilon_0$ being the free space electric constant and c the speed of light. In SLG we can take $\sigma_{2d,dc}= n\mu e$, where n is the number of charge carriers\cite{Geim_nmat}. Note that for n$\sim$0, $\sigma_{2d,dc}$ does not go to zero, but assumes a constant value\cite{Geim_nmat} $\sigma_{dc,min}\sim 4e^{2}/h$, resulting in R$_s$$\sim$6k$\Omega$ for an ideal intrinsic SLG with T$\sim 97.7\%$. Thus, ideal intrinsic SLG, would beat the best ITO only in terms of T, not R$_s$. However, samples deposited on substrates, or in thin films, or embedded in polymers are never intrinsic. Exfoliated SLG has typically n$\geq10^{12}cm^{-2}$ (see e.g. Ref. \onlinecite{Casi_APL}), and much smaller R$_s$.  The range of T and $R_{s}$ that can be realistically achieved for graphene layers of varying thickness can be estimated taking n=10$^{12}$-10$^{13}cm^{-2}$ and $\mu$=10$^{3}$-2$\times$10$^{4}$cm$^{2}$/Vs, as typical for CVD grown films. Figs. \ref{Figure_2}b,c show that graphene can achieve the same $R_{s}$ as ITO, ZnO-Ag-ZnO\cite{Sahu_ASS}, TiO$_2$/Ag/TiO$_2$ and SWNTs with a similar or even higher T. Fig. \ref{Figure_2}c plots T versus $R_s$ for ITO\cite {Peumans_NL}, Ag nanowires\cite {Peumans_NL}, SWNTs\cite{Geng_JACS} and the best graphene-based transparent conductive film (TCFs) reported to date\cite{Bae_arxiv}, again showing that the latter is superior. For instance, taking n=3.4$\times$10$^{12}$cm$^{-2}$ and $\mu$=2$\times$10$^{4}$cm$^{2}$/Vs, we get T=90$\%$ and $R_{s}=20\Omega/\Box$. Note that Eq.\ref{equ_13} is intended as guideline for TCF design and optimization, not as a statement on the transport physics of graphene. For TCF design empirical expressions of $\sigma_{2d,dc}$ as a function of carrier concentration and doping are enough, whatever the origin and precise quantification of the minimal conductivity, and of the dependence of R$_s$ on doping, defects, electron-hole puddles, etc.

Different strategies were explored to prepare graphene-based TCFs (GTCFs): spraying\cite{Gilje_Nanoletters}, dip\cite{Wang_NL} and spin coating\cite{Becerril_ACSnano}, vacuum filtration\cite{Eda_nnano}, roll-to-roll processing\cite{Bae_arxiv}.

Significant progress was made since the first attempts to produce GO based TCFs (GOTCFs). Since GO is insulating, it must be reduced to improve $R_{s}$\cite{Stankovich_Nature}. Ref. \onlinecite{Gilje_Nanoletters} decreased $R_{s}$ from $40G\Omega/\Box$ to $4M\Omega/\Box$ following reduction with dimethylhydrazine. Graphitization\cite{Becerril_ACSnano}, hydrazine exposure and low-temperature annealing\cite{Eda_nnano}, or high-temperature vacuum annealing\cite{WuACSnano} further decreased $R_{s}$, down to $800\Omega/\Box$ for T=82\%\cite{WuACSnano}.

Dispersions of graphite intercalated compounds (GIC)\cite{Biswas_NL} and hybrid nanocomposites (GO sheets mixed with silica sols or CNTs\cite{Tung_NL}) were also attempted, with a minimum $R_{s}=240\Omega/\Box$ for T=$86\%$\cite{Tung_NL}. Graphene films produced by chemical synthesis, currently show $R_{s}=1.6k\Omega/\Box$ for T=$55\%$\cite{Wang_Angwandte}.

Ref. \onlinecite{Blake_nanoletters} reported, thus far, the best GTCF from LPE of graphite. This was fabricated by vacuum filtration, followed by annealing, achieving R$_s$=$ 5k\Omega/\Box$; T$\sim 90\%$. The high $R_{s}$ is most likely due to the small flake size, and lack of percolation\cite{Blake_nanoletters,Green_NL}. The role of percolation can be seen in Ref. \onlinecite{Green_NL}, where $R_{s}$ and T went from $6k\Omega/\Box$;$\sim75\%$; to $2k\Omega/\Box$,$\sim77\%$ increasing flake size.

A key strategy to improve performance is stable chemical doping. Ref. \onlinecite{Blake_nanoletters} prepared GTCFs, produced by MC, with T$\sim98\%$;$R_{s}$=$400\Omega/\Box$, exploiting a layer of polyvinyl alcohol (PVA) to induce n-type doping. Ref. \onlinecite{Bae_arxiv} achieved R$_{s}$$\sim30\Omega/\Box$;T $\sim 90\%$ by nitric acid treatment of GTCFs derived from CVD grown flakes, one order of magnitude lower in terms of $R_{s}$ than previous GTCFs from wet transfer of CVD films\cite{Kim_nature}.

Figure \ref{Figure_2}d is an overview of current GTCFs and GOTCFs. It shows that GTCFs derived from CVD flakes, combined with doping, could outperform ITO, metal wires and SWNTs. Note that GTCFs and GOTCFs produced by other methods, such as LPE, albeit presently with higher $R_{s}$ at T=90$\%$, have already been tested in organic light emitters\cite{WuACSnano,Matyba_Acsnano} and solar cells\cite{Wang_NL,Liu_AM}. These are a cheaper and easier scalable alternative to MC or CVD films, and need be considered in applications where cost reduction is crucial.
\begin{figure*}
 \centerline{\includegraphics[width=170mm]{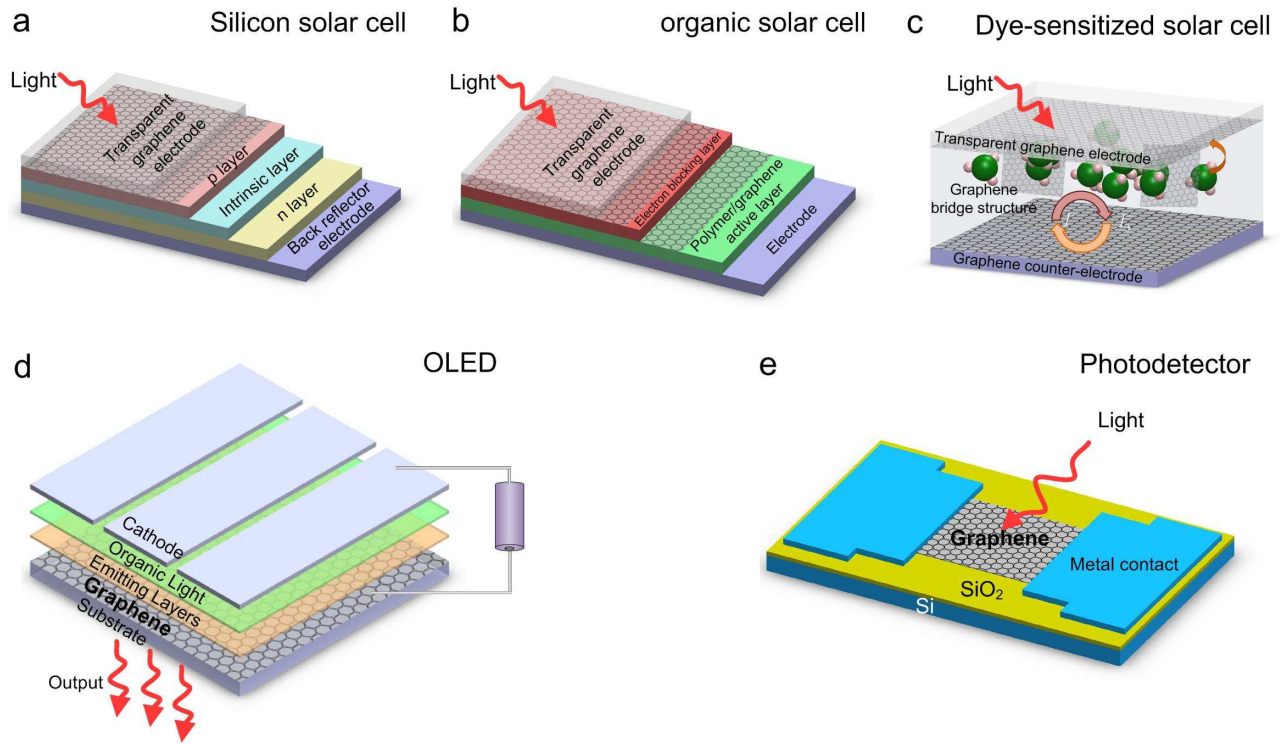}}
  \caption{\label{Figure_3} Schematic representation of (a) inorganic,(b) organic, and (c) dye-sensitized solar cell, (d) Organic Light Emitting Device (OLED), (e) photodetector.}
  \end{figure*}
\subsection{\label{OS} Photovoltaic devices}
A photovoltaic (PV) cell converts light to electricity\cite{Chapin_1954}. I$_{SC}$ is the maximum short circuit current, while V$_{OC}$ is the maximum open circuit voltage. The fill factor (FF) is defined as: FF=(V$_{max}$$\times $I$_{max}$)/(V$_{OC}$$\times$I$_{SC}$), where $V_{max}$ and $I_{max}$ are maximum voltage and current. The energy conversion efficiency is $\eta$=P$_{max}$/P$_{inc}$, where $P_{max}= V_{OC}\times I_{SC}\times FF $ and $P_{inc}$ is the incident power. The fraction of absorbed photons converted to current defines the internal photocurrent efficiency (IPCE).

Current PV technology is dominated by Si cells\cite{Chapin_1954}, with $\eta$ up to$\sim$25$\%$ \cite{Green_2001}. Organic photovoltaic cells (OPVs) rely on polymers for light absorption and charge transport\cite{Hoppe_04}. They can be manufactured economically compared to Si-cells, for example by a roll-to-roll process\cite{Kreebs_09}, even though they have lower $\eta$. An OPV consists of a TC, a photoactive layer and the electrode\cite{Hoppe_04}. Dye-sensitized solar cells (DSSCs) use a liquid electrolyte as a charge transport medium\cite{Gratzel_Nature}. A DSSC consists of a high porosity nanocrystalline photoanode, comprising TiO$_{2}$ and dye molecules, both deposited on a TC\cite{Gratzel_Nature}. When illuminated, the dye molecules capture the incident photon generating electron/holes pairs. The electrons are injected into the conduction band of the TiO$_{2}$ and transported to the counter electrode, the cathode\cite{Gratzel_Nature}. Regeneration of dye molecules is accomplished by capturing electrons from a liquid electrolyte. Presently, ITO is the most common material used both as photoanode and cathode, the latter with a Pt coating.

Graphene can fulfill multiple functions in PV devices: 1) TC window, 2) photoactive material, 3) channel for charge transport, 4) catalyst.

GTCFs can be used as window electrodes in inorganic (Fig. \ref{Figure_3}a), organic (Fig.\ref{Figure_3}b), and DSSCs devices (Fig. \ref{Figure_3}c). Ref. \onlinecite{Wang_Angwandte} used GTCFs produced by chemical synthesis, reporting $\eta\sim0.3\%$. Higher $\eta\sim0.4\%$ was achieved using reduced GO\cite{Wu_APL}, having $R_{s}=1.6k\Omega/\Box$\cite{Wu_APL} instead of $R_{s}=5k\Omega/\Box$\cite{Wang_Angwandte}, despite a lower T ($55\%$ instead of $80\%$). Ref. \onlinecite{De_Arco} achieved better performance ($\eta \sim1.2\%$) using CVD graphene as TCs with $R_{s}=230\Omega/\Box$ and T=72$\%$. Further optimization is certainly possible, considering the performance of the best GTCF to date\cite{Bae_arxiv}.

GO dispersions were also used in bulk hetero-junction PV devices, as electron-acceptors with poly(3-hexylthiophene)
and poly(3-octylthiophene) as donors, with $\eta \sim 1.4\%$\cite{Liu_AM}. Ref. \onlinecite{Yong_Small} claims that $\eta>12\%$ should be possible with graphene as photoactive material.

Graphene can cover an even larger number of functions in DSSCs. Ref. \onlinecite{Wang_NL} reported a solid-state DSSC based on spiro-OMeTAD1 (as hole transport material) and porous TiO$_{2}$ (for electron transport) using a GTCF anode, with $\eta \sim0.26\%$. Graphene can be incorporated into the nanostructured TiO$_{2}$ photoanode to enhance the charge transport rate, preventing recombination, thus improving the IPCE\cite{Yang_acsnano}. Ref. \onlinecite{Yang_acsnano} used graphene as TiO$_{2}$ bridge, achieving faster electron transport and lower recombination, leading to $\eta\sim7\%$, higher than conventional nanocrystalline TiO$_{2}$ photoanodes\cite{Yang_acsnano}. Another option is to use graphene to substitute the Pt counter electrode, due to its high specific surface area. An hybrid poly(3,4-ethylenedioxythiophene (PEDOT):poly-(styrenesulfonate) (PSS)/GO composite was used as counter electrode, getting $\eta$=4.5$\%$, comparable to 6.3$\%$, for a Pt counter electrode tested under the same conditions\cite{Hong_el_comm}, but now with a cheaper material.

\subsection{\label{LED} Light-emitting devices}
Organic light-emitting diodes (OLED) have an electroluminescent layer between two charge-injecting electrodes, at least one of which transparent\cite{Friend_Nature}. In OLED, holes are injected into the highest occupied molecular orbital (HOMO) of the polymer from the anode, while electrons are injected into the lowest unoccupied molecular orbital (LUMO) from the cathode. For efficiently injection, the anode and cathode work functions should match the HOMO and LUMO of the light-emitting polymer\cite{Friend_Nature}. Due to the image quality, low power consumption and ultra-thin device structure, OLEDs find applications in ultra-thin televisions and other display screens, such as computer monitors, digital cameras and mobile phones. Traditionally, ITO, having a work function 4.4-4.5eV, is used as TCF. However, besides cost issues, ITO is brittle and limited as flexible substrate\cite{Granqvist_07}. Also, In tends to diffuse into the active OLED layers, leading to a reduction of performance over time\cite{Hamberg_JAP_1986}. Thus, there is a need for alternative TCFs with optical and electrical performance similar to ITO, but without its drawbacks. Graphene has a work function of 4.5eV\cite{Giovannetti_PRL}, similar to ITO. This, combined with its promise as flexible and cheap TCF, makes it an ideal candidate as OLED anode, Fig. \ref{Figure_3}d, eliminating at the same time the issues related to In diffusion. GTCFs anodes enabled out-coupling efficiency comparable to ITO\cite{WuACSnano}. Considering that $R_{s}$ and T were $ 800\Omega/\Box$ and 82$\%$ at 550nm\cite{WuACSnano}, it is reasonable to expect further optimization will improve performance.

Ref. \onlinecite{Matyba_Acsnano} used a GOTCF in a light-emitting electrochemical cell (LEC). This is a device, similar to an OLED, where the light-emitting polymer is blended with an electrolyte\cite{Pei_Nat_Mat}. The mobile ions in the electrolyte rearrange when a potential is applied between the electrodes, forming high charge-density layers at each electrode interface, allowing efficient and balanced injection of electrons and holes, regardless of the work function of the electrodes\cite{Pei_Nat_Mat}. Usually, LECs have at least one metal electrode. Electrochemical side-reactions, involving the electrode materials, can cause problems in terms of operational lifetime and efficiency\cite{Matyba_Acsnano}. This also hinders the development of flexible LEC devices. Graphene is the ideal material to overcome these problems. Ref. \onlinecite{Matyba_Acsnano} demonstrated a LEC based solely on dispersion-processable carbon-based materials, paving the way to totally organic low-voltage, inexpensive, and efficient LEDs.

\subsection{\label{PD}Photodetectors}
Photodetectors (PDs) measure photon flux or optical power by converting the absorbed photon energy into electrical current. They are widely used in a range of common devices\cite{Saleh_book}, such as remote controls, TVs, DVD players, etc. Most PDs exploit the internal photo-effect, where the absorption of photons results in carriers excited from the valence to the conduction band, outputting an electric current. The spectral bandwidth is typically limited by the material absorption\cite{Saleh_book}. For example, PDs based on IV and III-V semiconductors suffer from the so-called "long-wavelength limit", as these become transparent when the incident energy is smaller than the bandgap\cite{Saleh_book}. Graphene absorbs from UV to THz\cite{Nair_Science,Sun_graphene,Dawlaty_APL_2,Wright_PRL}. As a result, graphene based PDs (GPDs) (see Fig. \ref{Figure_3}e) could work over a much broader wavelength range. The response time is ruled by the carrier mobility\cite{Saleh_book}. Graphene has huge mobilities. Hence, GPDs can be ultrafast.

The photoelectrical response of graphene has been widely investigated, both experimentally and theoretically\cite{Vasko_PRB_08,Park_2009_NL,Xia_NL_2009,Xia_nnano,Mueller_NP_10}. Photoelectrical response at 0.514, 0.633, 1.5 and 2.4$\mu$m was reported\cite{Mueller_NP_10}. Much broader spectral detection is expected due to the graphene ultrawideband absorption. Ref. \onlinecite{Xia_nnano} demonstrated a GPD with photo-response up to 40GHz. The GPD operation bandwidth is mainly limited by the time constant resulting from the device resistance R and capacitance C. Ref. \onlinecite{Xia_nnano}, reported a RC limited bandwidth$\sim640$GHz, comparable to traditional PDs\cite{Kang_NP_09}. However, the maximum possible operation bandwidth of PDs is typically restricted by the transit time, the finite duration of the photo-generated current \cite{Saleh_book}. The transit time limited bandwidth of a GPD could be well over 1500GHz \cite{Xia_nnano}, surpassing state-of-the-art PDs.

Although external electric field can produce efficient photocurrent generation, with $>30\%$ electron-hole separation efficiency\cite{Park_2009_NL}, zero source-drain bias and dark current operations could be achieved by using the internal electric field formed near the metal electrode-graphene interfaces\cite{Xia_nnano,Mueller_NP_10}. The small effective area of the internal electric field could decrease the detection efficiency\cite{Xia_nnano,Mueller_NP_10}, since most of the generated electron-hole pairs would be out of the electric field, thus recombining, rather than being separated. The IPCE ($15-30\%$)\cite{Xia_NL_2009,Park_2009_NL} and external responsivity (generated electric current for given input optical power) of 6.1mA/W\cite{Mueller_NP_10} so far reported for GPDs are relatively low compared to current PDs\cite{Saleh_book}. This is mainly due to limited optical absorption when only one SLG is used, short photo-carrier lifetime, and small effective photo-detection area ($\sim 200nm$ in Ref. \onlinecite{Xia_nnano}).

The photo-thermoelectric effect, which exploits the conversion of photon energy into heat and then electric signal\cite{Saleh_book}, may play an important role in photocurrent generation in graphene devices\cite{Park_2009_NL,Xu_NL_10}. Thus photo-thermoelectric GPDs may be possible.
\begin{figure}
\centerline{\includegraphics[width=70mm]{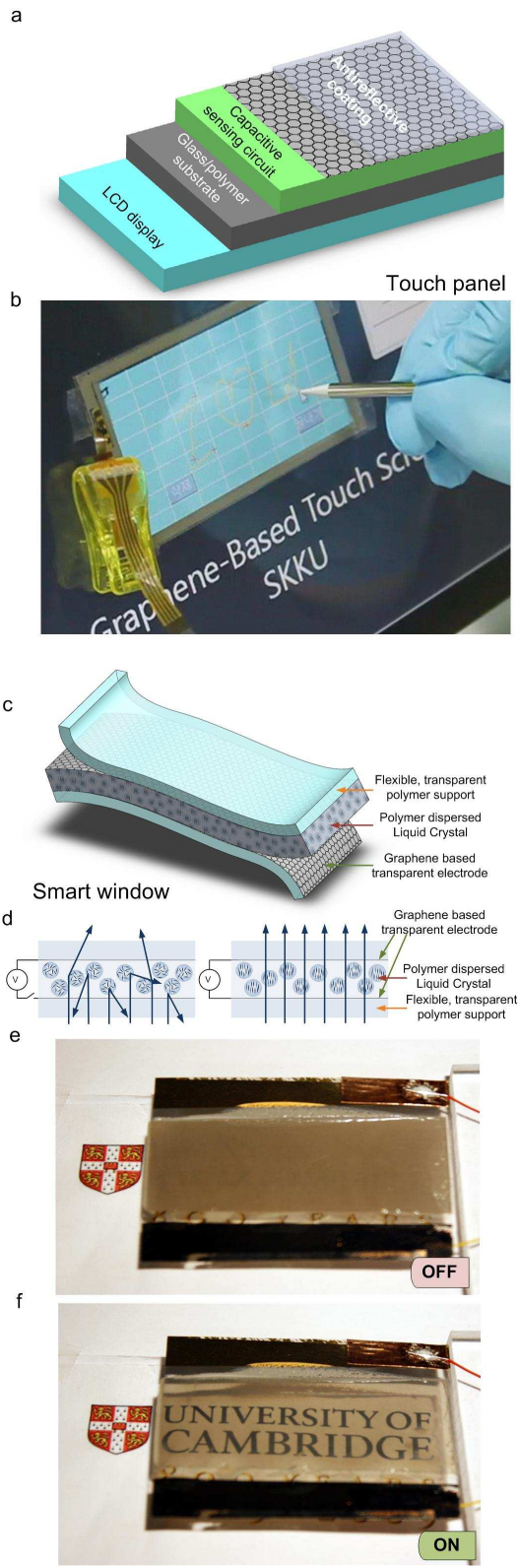}}
\caption{\label{Figure_4}a) Schematic capacitive touch panel. b) Resistive graphene-based touch screen (adapted from Ref. \onlinecite{Bae_arxiv}). c) Schematic of PDLC smart window using a GTCF. d) With no voltage, the LC molecules are not aligned, making the window opaque. e,f) Graphene/Nanotube-based smart window in off/on state.}
\end{figure}
\subsection{\label{TP} Touch screens}
Touch screens are visual outputs that can detect the presence and location of a touch, by a finger or other objects, such as a pen, within the display area, thereby permitting the physical interaction with what shown on the display itself\cite{Maeda_99}. Touch panels are currently used in a wide range of applications, such as cellular phones and digital cameras, and where keyboard and mouse do not allow a satisfactory, intuitive, quick, or accurate interaction by the user with the display content.

Resistive and capacitive (see Fig. \ref{Figure_4}a) touch panels are the most common. A resistive touch panel comprises a conductive substrate, a liquid crystal device (LCD) frontpanel, and a TCF\cite{Maeda_99}. When pressed by a finger or pen, the frontpanel film comes into contact with the bottom TC and the coordinates of the contact point are calculated on the basis of their resistance values. There are two categories of resistive touch screens: matrix and analogue\cite{Maeda_99}. The matrix has striped electrodes, while the analogue has a non-patterned TC electrode, thus lower production costs. The TC requirements for resistive screens are $R_{s} \sim 500-2000\Omega/\Box$ and T$>$90$\%$ at 550nm\cite{Maeda_99}. Favorable mechanical properties, including brittleness and wear resistance, high chemical durability, no toxicity, and low production costs are also important. Cost, brittleness, wear resistance, and chemical durability are the main limitations of ITO\cite{Hamberg_JAP_1986,Granqvist_07}. Thus for resistive touch-screens there is an effort to find an alternative TC. Indeed, ITO cannot withstand repeated flexing and poking involved with this type of touch screens without cracking or deteriorating.

GTCFs can satisfy the requirements for resistive touch screens in terms of T and $R_{s}$, when combined with large area uniformity. Ref. \onlinecite{Bae_arxiv} recently reported a graphene-based touch panel display by screen-printing a CVD grown sample, Fig. \ref{Figure_4}b. Considering the $R_{s}$ and T required by analogue resistive screens, GTCF or GOTCF produced via LPE also offer a viable alternative, and further cost reduction.

Capacitive touch screens are emerging as the high-end version, especially since the launch of Apple's iPhone. These consist of an insulator such as glass, coated with ITO\cite{Maeda_99}. As the human body is also a conductor, touching the surface of the screen results in a electrostatic field distortion, measurable as a change in capacitance. Although capacitive touch screens do not work by poking with a pen, thus mechanical stresses are lower with respect to resistive ones, the use of GTCFs can improve performance and reduce costs.

\subsection{\label{ESSW} Flexible smart windows}
Polymer dispersed liquid crystal devices (PDLC) or similar structures, generally known as 'smart windows', were introduced in the early 80s\cite{craighead_82}. These consist of thin films of optically transparent polymers with micron-sized LC droplets contained within pores of the polymer. Light passing through the LC/polymer is strongly forward scattered, producing a milky film\cite{drzaic_86,doane_88}. If the LC ordinary refractive index is close to that of the host polymer, the application of an electric field results in a transparent state\cite{sheraw_02}. The ability of switching from translucent to opaque makes them attractive in many applications, e.g. where privacy at certain times is highly desirable. There are other potential applications of PDLC in flexible displays due to their compatibility with flexible substrates and wide viewing angle\cite{sheraw_02}. For example, Ref. \onlinecite{sheraw_02} demonstrated an organic thin film transistor driven flexible display with each individual pixel controlled by an addressable PDLC matrix. Conventionally, ITO on glass is used as TCF to apply the electric field across the PDLC. However, one of the reasons behind the limited market penetration of smart windows is the significant ITO cost. Furthermore, flexibility is hindered when using ITO, reducing potential applications, such as PDLC flexible displays\cite{sheraw_02}. For transparent or colored/tinted smart windows, the generally required T and $R_{s}$ range from 60 to 90$\%$ and above and 100 to 1k$\Omega/\Box$ depending on production cost, application and manufacturer. In addition to flexibility, the electrodes need to be as large as the window itself and must have long term physical and chemical stability, as well as being compatible with the roll to roll PDLC production process.

All these deficiencies of ITO electrodes can be overcome by GTCFs. Figures \ref{Figure_4}c,d shows the working principle and Fig. \ref{Figure_4}e,f a prototype of a flexible smart window with PET used as substrate.
\begin{figure}
 \centerline{\includegraphics[width=80mm]{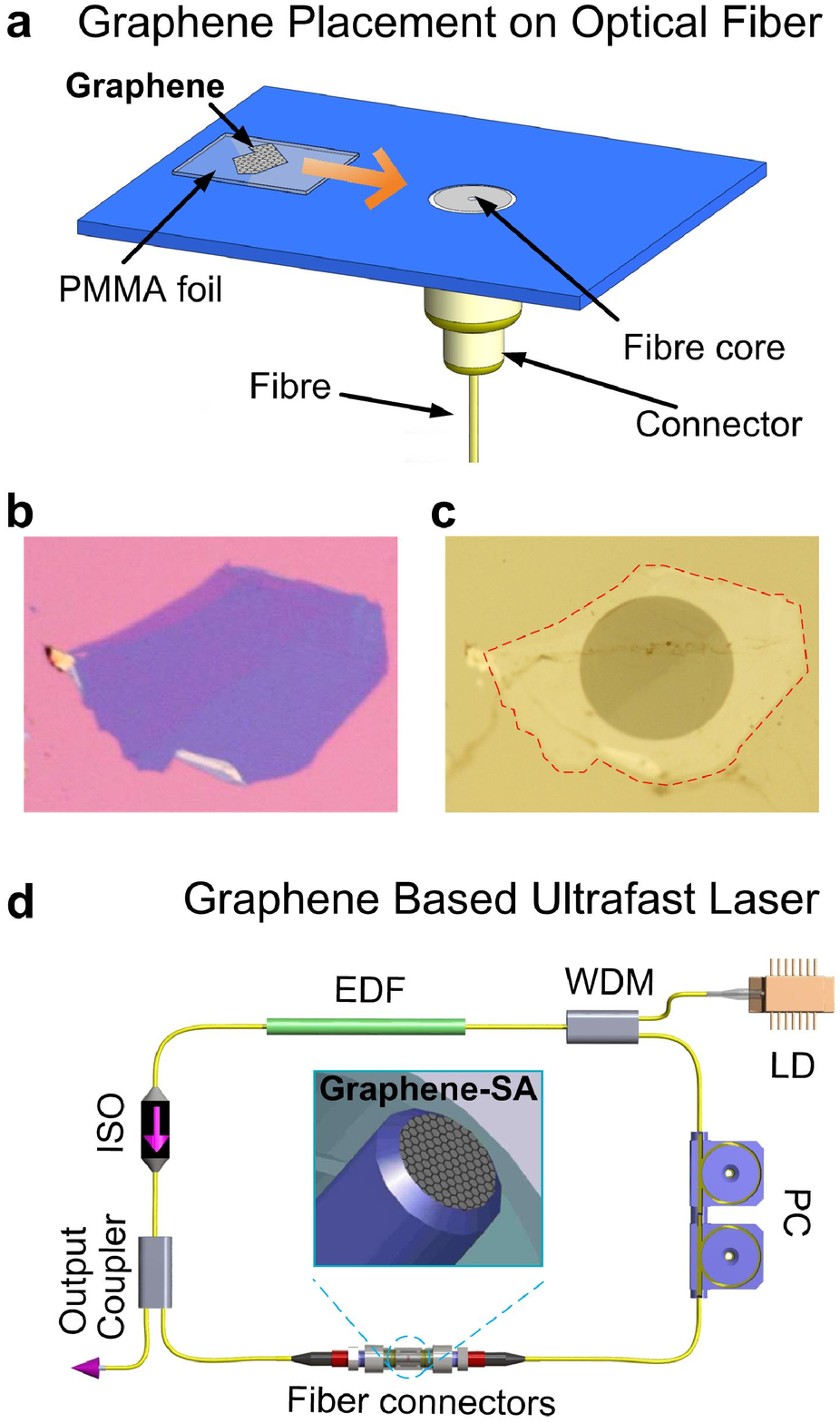}}
  \caption{\label{Figure_5} (a) an optical fibre is mounted onto a holder. Once detached from the original substrate, a polymer/graphene membrane is slid and aligned with the fibre core. b) Flake originally deposited on $SiO_{2}/Si$. c) The same flake after deterministic placement. d) graphene-mode locked ultrafast laser: a graphene saturable absorber is placed between two fiber connectors. An Erbium doped fibre (EDF) is the gain medium, pumped by a laser diode (LD) via a wavelength-division-multiplexer (WDM). An isolator (ISO) maintains unidirectional operation. A polarization controller (PC) optimizes mode-locking.}
\end{figure}
\begin{figure*}
\centerline{\includegraphics[width=140mm]{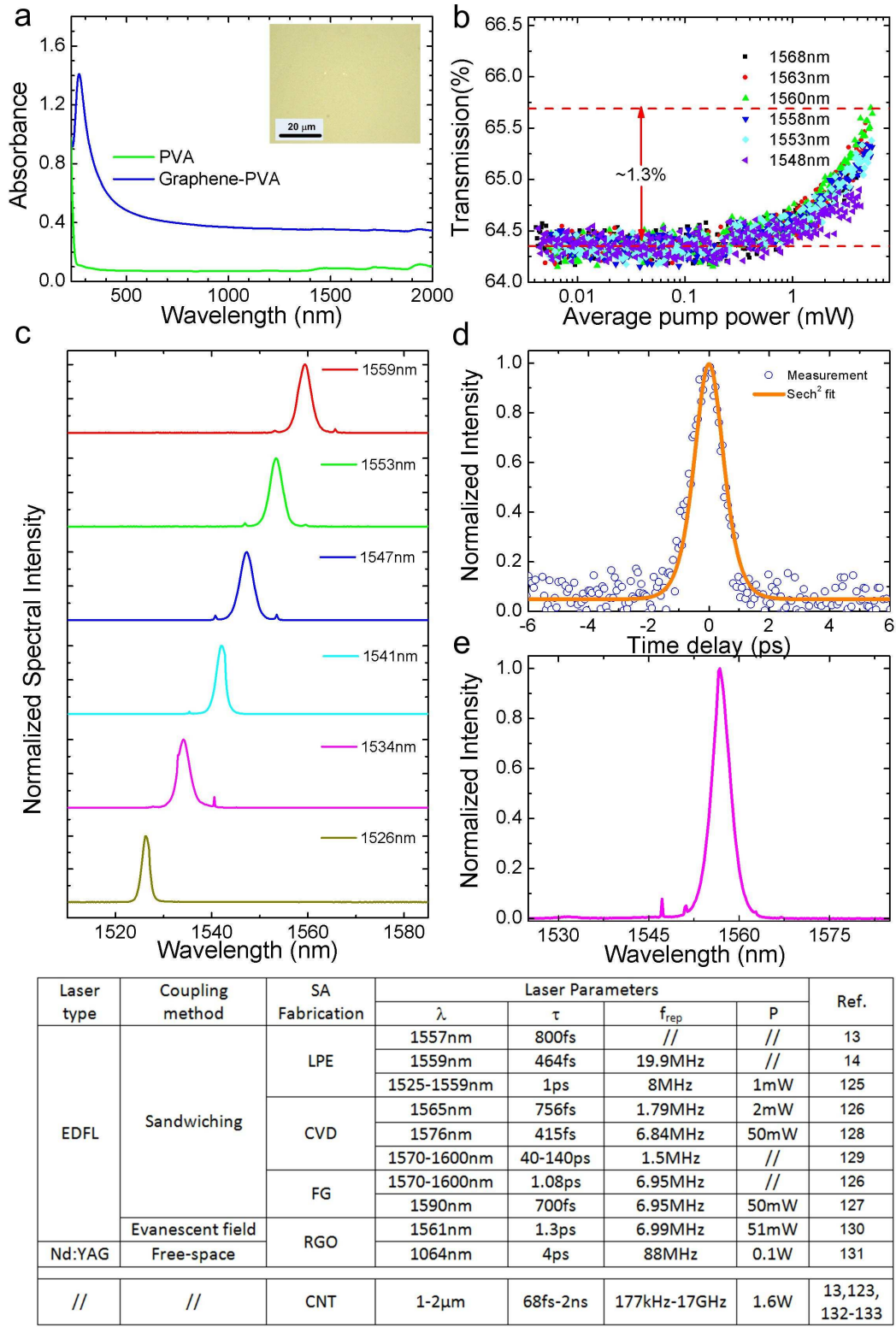}}
\caption{\label{Figure_6} a) Absorption of graphene-PVA composite and reference PVA. Inset: micrograph of the composite.(b) Typical transmission as a function of pump power at six wavelengths. T increases with power.(c) Tuneable ($>$30nm) fiber laser mode-locked by graphene. (d) Autocorrelation (AC) trace and (e) spectrum of output pulses of a GO-mode-locked laser, with a$\sim$743fs pulse duration. Table: GSA mode-locked laser performances.EDFL: Erbium-doped fiber laser; FG: Functionalized graphene; Nd:YAG SSL: Neodymium doped ytterbium aluminum garnet solid-state laser.}
\end{figure*}
\subsection{\label{SA} Saturable absorbers and ultrafast lasers}
Materials with non-linear optical and electro-optical properties are needed in most photonic applications\cite{Boyd_book,Hasan_AM}. Laser sources producing nano- to sub-picosecond pulses are a major component in the portfolio of leading laser manufacturers. Thus far, solid-state lasers are the short-pulse source of choice, being deployed in a variety of applications ranging from basic research to material processing, from eye surgery to printed circuit board manufacturing, from metrology to trimming of electronic components such as resistors and capacitors. Regardless of wavelength, the majority of ultrafast laser systems employ a mode-locking technique, whereby a non-linear optical element, called saturable absorber (SA), turns the continuous wave output into a train of ultrafast optical pulses\cite{Keller_nature_03}. The key requirements for non-linear materials are fast response time, strong non-linearity, broad wavelength range, low optical loss, high power handling, low power consumption, low cost and ease of integration into an optical system. Currently, the dominant technology is based on semiconductor saturable absorber mirrors (SESAMs)\cite{Keller_nature_03}. However, these have a narrow tuning range, and require complex fabrication and packaging \cite{Keller_nature_03,Hasan_AM}. A simple, cost-effective alternative is to use SWNTs\cite{Hasan_AM,Wang_Tunable}, where the diameter controls the gap, thus the operating wavelength. Broadband tunability is possible using SWNTs with a wide diameter distribution\cite{Hasan_AM,Wang_Tunable}. However, when operating at a particular wavelength, SWNTs not in resonance are not used and contribute unwanted losses.

As discussed in Sect.\ref{sat_ab}, the linear dispersion of the Dirac electrons in graphene offers an ideal solution: for any excitation there is always an electron-hole pair in resonance. The ultrafast carrier dynamics\cite{Breusing_PRL,SUN_PRL}, combined with large absorption and Pauli blocking make graphene an ideal ultra-broadband, fast SA. Compared to SESAMs and SWNTs, there is no need of bandgap engineering or chirality/diameter control.

Thus far, graphene-polymer composites\cite{Hasan_AM,Sun_graphene,Sun_Tunable,Bao_AFM_2010,Zhang_APL_gra}, CVD grown films\cite{Zhang_OE_gra,Zhang_apl_10_tun}, functionalized graphene (e.g. GO bonded with poly[(m-phenylenevinylene)-co-(2,5-dioctoxy-p-phenylenevinylene)])\cite{Bao_AFM_2010} and reduced GO flakes\cite{Song_APL_2010,Tan_APL_2010} were used for ultrafast lasers. Graphene-polymer composites are scalable and, more importantly, allow easy integration into a range of photonic systems\cite{Hasan_AM,Sun_graphene,Sun_Tunable}. Another route for graphene integration is via deterministic placement in a pre-defined position on a substrate of choice, e.g. a fiber core or cavity mirrors. Fig.\ref{Figure_5}a shows such transfer of a flake onto an optical fibre core. This is achieved by exploiting a water layer between the poly(methyl methacrylate)(PMMA)/graphene foil and the optical fibre, which enables the PMMA to move. Graphene integration is finally obtained after precise alignment to the fibre core by a micro-manipulator (Fig. \ref{Figure_5} b) and dissolution of the PMMA layer (Fig.\ref{Figure_5}c).

A typical absorption spectrum is shown in Fig.\ref{Figure_6}a \cite{Hasan_AM,Sun_graphene,Sun_Tunable}. This is featureless, save a characteristic UV plasmon peak, while the host polymer only contributes a small background for longer wavelengths. Fig.\ref{Figure_6}b plots T as a function of average pump power at six wavelengths. Saturable absorption is evident by the T increase with power at all wavelengths.

Various strategies have been proposed to integrate graphene saturable absorbers (GSAs) in laser cavities for ultrafast pulse generation. The most common is to sandwich a GSA between two fibre connectors with a fibre adapter, as schematized in Fig.\ref{Figure_5}c \cite{Hasan_AM,Sun_graphene,Sun_Tunable}. Graphene on a side-polished fiber was also reported, aimed at high power generation by evanescent field interaction\cite{Song_APL_2010}. A quartz substrate coated with graphene was used for free-space solid-state lasers\cite{Tan_APL_2010}.

The most common wavelength of generated ultrafast pulses so far is$\sim1.5\mu m$, since this is the standard in optical telecommunications, not because GSAs have any preference for a particular wavelength. A solid-state laser mode-locked by graphene was reported at$\sim 1 \mu m$\cite{Tan_APL_2010}. Fig.\ref{Figure_6}c shows a GSA-mode locked EDF laser tuneable from 1526 to 1559nm\cite{Sun_Tunable}, with the tuning range mainly limited by the tunable filter, not the GSA. Fig.\ref{Figure_6}d,e show the pulse from a GO based saturable absorber (GOSA). The possibility of tuning the GSA properties by functionalization and using different layers or composite concentrations, offers considerable design freedom. Fig.\ref{Figure_6} has a table comparing the performance of graphene-based ultrafast lasers and the main CNT-based devices\cite{Scardaci_am_2008,Sun_apl_2009}.

\subsection{\label{OL} Optical limiters}
Optical limiters (OL) are devices with high transmittance for low incident light intensity, and vice-versa for high intensity\cite{Bass_book}. There is a great interest in OL mainly for optical sensors and human eye protection\cite{Bass_book}, as damage can happen when the intensity exceeds a threshold\cite{Bass_book}. Passive OL using a nonlinear optical material, have potential for simplicity, compactness and low cost\cite{Bass_book}. However, so far there is no demonstration of passive OL able to protect eyes and other common sensors over the entire visible and NIR range\cite{Bass_book}. Typical OL materials include semiconductors (e.g. ZnSe, InSb), organic molecules (e.g. phthalocyanines), LC and carbon based composites (e.g. carbon-black dispersions, CNTs, fullerenes)\cite{Bass_book,Wang_advmater}. Fullerenes and their derivatives\cite{Tutt_Nat_92,Wang_jmc_09} and CNTs dispersions\cite{Wang_jmc_09} have good OL performance, in particular for ns pulses at 532 and 1064nm\cite{Wang_jmc_09}.

In graphene-based OL the absorbed light energy converts into heat, creating bubbles and microplasmas\cite{Wang_advmater}, resulting in reduced transmission. Graphene dispersions can be used as wideband OL covering visible and NIR. Broad OL (from 532 to 1064nm) of LPE graphene was reported for ns pulses\cite{Wang_advmater}. It was shown that funtionalized graphene dispersions could outperform $C_{60}$\cite{Xu_AM_2009}.

\subsection{\label{OFC} Optical Frequency converters}
Optical frequency converters are used to expand the wavelength accessibility of lasers (e.g. frequency doubling, parametric amplification and oscillation, and four-wave mixing (FWM))\cite{Bass_book}. Calculations suggest that nonlinear frequency generation in graphene (e.g. harmonics of input light) should be possible for sufficiently high external electric field ($>100V/cm$)\cite{Mikhailov_EPL_2007}. Second harmonic generation (SHG) from a 150fs laser at 800nm was reported for a graphene film\cite{Dean_APL_2009}. In addition, FWM, generating NIR wavelength-tunable light was demonstrated using SLG and FLG\cite{Hendry_2009_FWM}. Graphene's third-order susceptibility was measured to be $|\chi^{3}| \sim 10^{-7} esu$\cite{Hendry_2009_FWM}, one order of magnitude larger than reported so far for CNTs \cite{Hendry_2009_FWM}. However, photon-counting electronics is typically needed to measure the output\cite{Dean_APL_2009}, indicating a low conversion efficiency. Other features of graphene, such as the possibility of tuning the nonlinearity by changing the number of layers\cite{Hendry_2009_FWM}, and wavelength-independent nonlinear susceptibility\cite{Hendry_2009_FWM}, still could be potentially used for various photonic applications (e.g. optical imaging \cite{Hendry_2009_FWM}).

\subsection{\label{Thz}THz devices}
THz radiation in the 0.3 to 10THz range ($30\mu m$ to $1 mm$), is attractive for biomedical imaging, security, remote sensing and spectroscopy\cite{Thz_book}. A large amount of unexplored territory for THz technology still remains mainly due to a lack of affordable and efficient THz sources and detectors\cite{Thz_book}. The frequency of graphene plasma waves\cite{Rana_2008,Ryzhii_JAP_07_1} lies in the THz range, as well as the gap of GNRs, and the BLG tunable band gap, making graphene appealing for THz generation and detection. Various THz sources were suggested, based on electrical\cite{Rana_2008,Ryzhii_JAP_07_1} or optical\cite{Rana_2008,Ryzhii_JAP_07_1} pump of graphene devices. Recent experimental observations of THz emission\cite{Sun_NL_10} and amplification\cite{Otsuji_10_THz} in optically pumped graphene show the feasibility of graphene-based THz generation. Twisted multilayers, retaining the electronic properties of SLG, could also be interesting for THz applications.

Graphene devices can be used for THz detection and frequency conversion. The possibility of tuning the electric and optical properties by external means (e.g. electric and magnetic field, optical pump), makes SLG and FLG suitable for IR and THz radiation manipulation as well. The possible devices include modulators, filters, switches, beam splitters and polarizers.

\section{\label{Conclusion}Perspective}
Graphene films and composites have attractive electronic and optical properties, making them ideal for photonics and optoelectronics. A number of applications are enabled by using graphene as a replacement for ITO and other transparent conductors. In many cases (e.g. touch-screens, OLEDs) this also adds fabrication flexibility, in addition to economic advantages. Current PDLC-based devices face major challenges due to fabrication costs associated with the requirement of large transparent electrodes. The move to a graphene-based technology could make them more viable. Novel graphene-based transparent electrodes on flexible substrates for solar cells can add value and operational flexibility, not possible with current transparent conductors and rigid glass substrates. Recent progresses on growth and dispersion processing of graphene have definitely made this material "come of age", encouraging industrial applications. Deterministic placement of graphene layers on arbitrary substrates, and multi-layers by individual assembly of monolayers at given angles are now possible.  Future efforts on nonlinear optical devices will focus on demonstrators at different wavelengths to fully exploit graphene's ultrawide broadband capability. Ultrafast and tunable lasers are a reality, with an ever growing number of groups entering this field. The combination of graphene photonics with plasmonics could enable a variety of advanced devices.

\section{\label{Conclusion}Acknowledgments}
We thank S. A. Awan, D.M. Basko, E. Lidorikis, A. Hartschuh, J. Coleman, A. Dyadyusha, T. Kulmala, A. Lombardo, D. Popa, G. Privitera, F. Torrisi, O. Trushkevych, F. Wang, T. Seyller,B. H. Hong, A.K. Geim, K.S. Novoselov for useful discussions. We acknowledge funding from EPSRC grants GR/S97613/01, EP/E500935/1, ERC grant NANOPOTS, a Royal Society Brian Mercer Award for Innovation, the Cambridge Integrated Knowledge Centre in Advanced Manufacturing Technology for Photonics and Electronics. F.B. acknowledges funding from a Newton International Fellowship and T.H. from King's College, Cambridge. A.C.F. is a Royal Society Wolfson Research Merit Award holder.

\end{document}